\documentclass[preprint,12pt]{elsarticle}

 \usepackage{graphicx}

\usepackage{amssymb}

\newcommand{\bbar}   {\bar b}
\newcommand{\tbar}   {\bar t}

\newcommand{\shat}    {{\hat s}}

\def\slashii#1{\setbox0=\hbox{$#1$}            
  \dimen0=\wd0                                 
  \setbox1=\hbox{\sl/} \dimen1=\wd1            
  \ifdim\dimen0>\dimen1                        
     \rlap{\hbox to \dimen0{\hfil\sl/\hfil}}   
     #1                                        
  \else                                        
     \rlap{\hbox to \dimen1{\hfil$#1$\hfil}}   
     \hbox{\sl/}                               
  \fi}

\journal{Acta Physics Polonica}

\begin{document}

\begin{frontmatter}



%
\title{Background to Higgs-boson searches from internal conversions 
of off-shell photons associated with  $Z/\gamma ^*$-boson production at the LHC$^{\star}$}


\author[1,2]{Anatoli Fedynitch}
\author[3]{Mieczyslaw Witold Krasny}
\author[4]{Wies{\l}aw P{\l}aczek}
\address[1]{CERN, CH-1211 Geneva 23, Switzerland.}
\address[2]{Karlsruher Institut f\"ur Technologie, Institut f\"ur Kernphysik, \\
Postfach 3640, 76021 Karlsruhe, Germany.}
\address[3]{Laboratoire de Physique Nucl\'eaire et des Hautes Energies, \\
          Universit\'e Pierre et Marie Curie -- Paris 6, Universit\'e Paris Diderot -- Paris 7, \\
          CNRS--IN2P3, 4 pl. Jussieu, 75005 Paris, France.}
\address[4]{Marian Smoluchowski Institute of Physics, Jagiellonian University,\\
         ul.\ {\L}ojasiewicza 11, 30-348 Krak\'ow, Poland.}
\begin{abstract}
This paper  presents the studies of the background  contribution to the $H \rightarrow 4l$ searches  originating  from 
the processes of off-shell (virtual) photon emissions and their conversions into lepton pairs 
accompanying  the production of  $Z/  \gamma ^*$-bosons at the LHC.
They extend the analyses of the irreducible background presented in  the ATLAS and CMS Higgs papers  
\cite{ATLAS_Higgs, CMS_Higgs}  
by taking into account the emissions of off-shell photons by parton showers.  
Including 
these effects
does not change significantly the Higgs-searches background level, 
provided that the transverse momentum of each of the final-state leptons is restricted to the range of $p_{T, l} > 7$ GeV. In the 
kinematical region extended towards lower lepton transverse momenta the parton-shower contribution becomes important.
A measurement method for pinning down the parton-shower effects has been proposed.
\end{abstract}


\end{frontmatter}

\vspace{2mm}
\footnoterule
\noindent
{\footnotesize
$^{\star}$The work is partly supported by the Programme of the French--Polish 
Cooperation between IN2P3 and COPIN no.\ 05-116,
and by the Polish National Centre of Science grant no.\ DEC-2011/03/B/ST2/00220.}


\section{Introduction}
\label{Introduction}

The Higgs boson discovery \cite{ATLAS_Higgs,CMS_Higgs},
is  the first  out of all the pivotal {\it experimental} particle physics discoveries which relies,
to such a large extent, on  the {\it theoretical} calculations of the Standard 
Model (SM)  background. 

Given its importance for the future experimental program in  high energy physics, it should be of utmost interest to keep 
exposing the theoretical background calculations to a broad spectrum of experimental and 
theoretical stress tests,  no matter how widely this discovery is acclaimed.

Majority of the SM background  processes leading to Higgs-like signatures 
have already been identified and extensively 
studied  experimentally, using the data extrapolation driven techniques,   
and theoretically, using the existing Monte Carlo generators of SM processes. 
The basic two questions which have motivated our studies presented in this 
and in two other papers \cite{DDYP1,DDYP2} are: 
\begin{enumerate}
\item Is the list of the background sources complete? 
\item Are  various approximations inherent in the  Monte Carlo generators, 
used in the determination of the experimentally irreducible background, 
controlled to the claimed precision?
\end{enumerate}

In our previous work \cite{DDYP1}, see also \cite{DDYP2}, we have concentrated our attention on
the double Drell--Yan process (DDYP),  considered to be negligible in the ATLAS and CMS analyses. 
We have demonstrated, using a  simplified model of DDYP,  the appearance 
of a  peak in the four-lepton invariant  mass, $m_{4l}$,  distribution which mimics the $125$ GeV 
Higgs signal in its $H \rightarrow ZZ^*$ decay channel.  
This  ``Higgs-like" peak is generated by the  interplay of a steeply falling $m_{4l}$ 
distribution  and the kinematical  threshold effect driven by  the experimental cuts on the outgoing leptons variables. 
The cuts influencing the background peak position are similar in the ATLAS and CMS analyses and are therefore reflected 
in similar peak positions within a $2$ GeV interval. The coincidence of the Higgs peak and the DDYP peak 
could be accidental. It is, however remarkable, that once  the absolute  normalisation 
of the DDYP contribution is fixed  using the LHC $W^+W^-$ cross-section  data \cite{DDYP2}, the DDYP process
could provide an alternative explanation of the Higgs-like excesses of the events both in the $H \rightarrow ZZ^*$ 
and in the $H \rightarrow WW^*$ channels.

The claim of  the ATLAS and CMS collaborations that DDYP can be neglected 
as a potentially alarming source of background was based on the 
assumption of uncorrelated: (1) longitudinal momentum, (2) transverse position, (3) flavour, (4) charge and (5) spin 
of the partons taking part in DDYP,  and on the  assumption  of the process independent value of $\sigma_{\rm eff}$,  
governing the strength of double-parton scattering (DPS) processes \cite{ATLAS_Higgs,CMS_Higgs}.
The above, in our view unjustified,  assumptions  lead  to a significant underestimation of the contribution 
of DDYP to the Higgs searches background. As we argued in \cite{DDYP1}, its contribution must be, given the lack of the 
adequate theoretical calculations,  determined experimentally, e.g.\ by using experimental methods proposed therein
and in \cite{KRASNY_S_dependence}.

\begin{figure}[h]
\begin{center}
\includegraphics[width=1.00\textwidth]{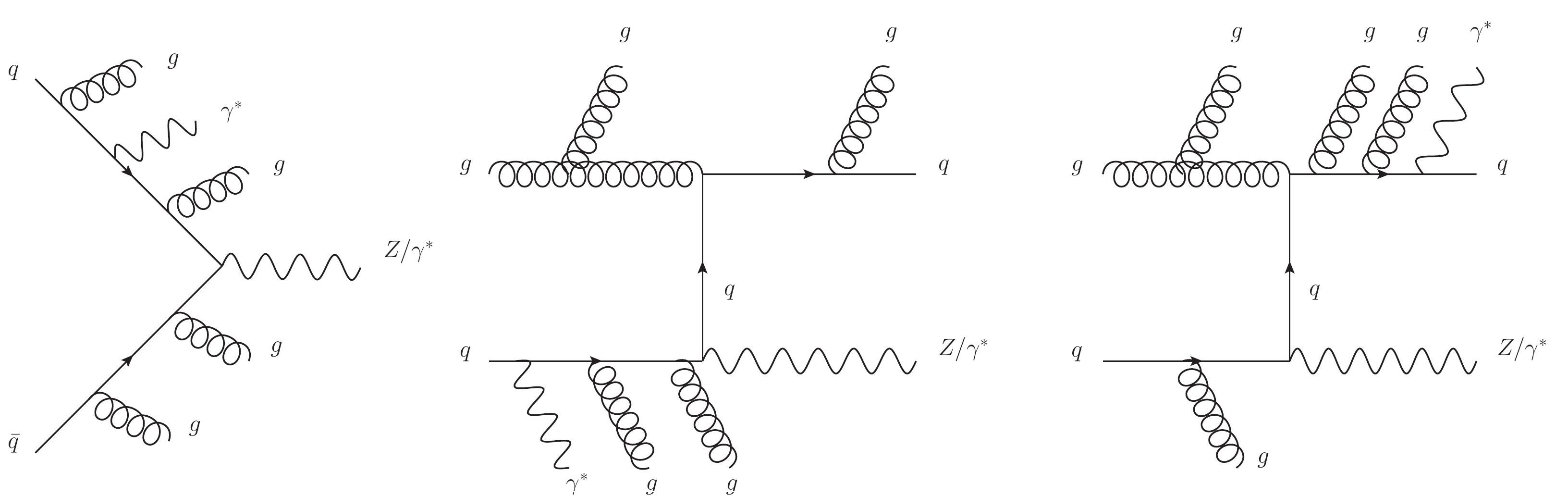} 
\end{center}
\caption{Some examples of the parton-shower diagrams for the inclusive  $Z / \gamma^*$ production
processes  contributing to the Higgs searches background. The final-state 
leptons are produced by the decays of the $Z / \gamma^*$ boson   and by the internal conversions of virtual photons, $\gamma ^*$.
The virtual photons can be emitted at an arbitrary stage of the parton shower.}
\label{PS-diagrams}
\end{figure}

In the current paper we study another potentially important  contribution to 
the Higgs searches background, which has not been entirely taken into account 
in the ATLAS and CMS papers \cite{ATLAS_Higgs,CMS_Higgs}, namely 
the processes of emission and subsequent internal conversion of off-shell (virtual) photons 
radiated by quarks and leptons participating in the production and leptonic decays of $Z / \gamma^*$ bosons. 
Examples of the Feynman diagrams corresponding to these processes are shown in  
Fig.~\ref{PS-diagrams}. 

The Matrix Element (ME) contribution to  these processes  -- for the case of {\it quark--antiquark annihilations}  -- 
was considered  in the ATLAS and the CMS analyses  \cite{ATLAS_Higgs, CMS_Higgs} and evaluated 
using the {\sf POWHEG-BOX/ZZ} next-to-leading-order 
(NLO) ME generator ({\sf POWHEG}) \cite{POWHEG}. The {\sf POWHEG} quark--antiquark annihilation 
contribution to the Higgs searches background in  $ZZ^*$ channel was found to be the dominant one   
\cite{ATLAS_Higgs, CMS_Higgs}.

\begin{figure}[h]
\begin{center}
\includegraphics[width=1.00\textwidth]{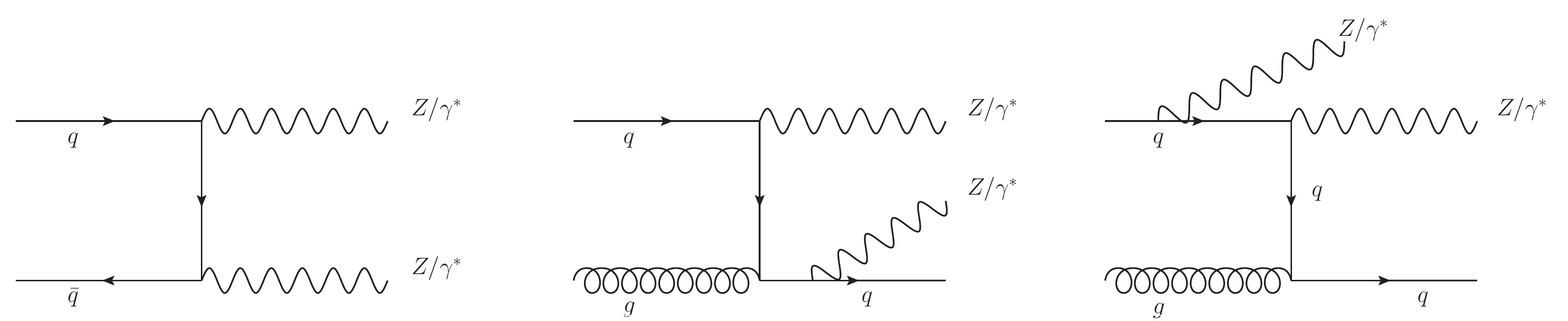} 
\end{center}
\caption{The lowest-order matrix element (ME) contributions to the processes of $Z / \gamma^*$ pair  production in quark--antiquark and quark--gluon scattering.}
\label{QG-diagrams}
\end{figure}

The {\em first} question  which, in our view,  must be addressed   is if the contribution of the {\it quark--gluon  scattering processes} 
can be neglected. These  processes are suppressed by an ${\cal O}(\alpha_s)$ factor. 
However, this suppression may be to a great extent neutralised by a large value of the ratio of the 
gluon to quark parton distribution functions (PDFs) in the $x$-domain which is pertinent to the Higgs signal extraction. 
The lowest-order quark--antiquark and the quark--gluon  scattering diagrams for $Z / \gamma^*$ pair production are shown in Fig.~\ref{QG-diagrams}.

 The {\em second} question is whether  the contribution of processes in which  
 virtual photons are emitted by {\it parton showers} (PS), both for the quark--antiquark 
 and quark--gluon collisions,  can be neglected for the 
 inclusive four-lepton analysis\footnote{In the fully inclusive searches both  the number 
 of lepton pairs and  the four-lepton phase space are not restricted. }. 
The strength of this contribution increases with decreasing off-shellness of 
virtual photons  proportionally  to
 $\ln ^2 (\shat/m^2_{\gamma ^*})$, where $\shat$  is the invariant mass of the four-lepton system 
 and $m_{\gamma ^*}$ is the virtual photon mass \cite{BvNB:1988}.  
 
In this paper we address the latter question, while the former one will be  addressed in a separate publication \cite{Anatoli_future}. 
 
The ``PS" question cannot, at present, be answered at NLO precision, because an appropriate NLO Monte Carlo generator 
producing interleaved QED and QCD parton showers, including the full set of the re-summed  $\alpha^2_{\rm QED}$ corrections 
to the inclusive production of vector bosons, does not exist. The studies presented 
in this paper  are based on the leading-order (LO) 
{\sf PYTHIA~8} event generator \cite{PYTHIA}, which has inferior quality with respect to {\sf POWHEG} in calculating 
the NLO ME contribution to the four-lepton final state. However, virtual photon emissions by the initial and 
final state quarks and leptons are included in all stages of the QCD/QED-interleaved parton shower.

The {\sf PYTHIA} PS description of the virtual photon emissions  has never, to our best knowledge, 
been tested in the kinematical domain discussed in this paper. Therefore,  an evaluation of its precision is 
the prerequisite for its subsequent use in the studies of the Higgs searches background. 
Such an evaluation is presented in Section \ref{calibration},
following the introductory discussion of the background sources to the Higgs searches in the 
four-lepton channel. This discussion is  presented
in Section \ref{background}.  In Section \ref{results} the results of our calculations are presented. 
The proposal of an experimental test  aiming to verify the importance of  the QED/QCD  PS 
contribution to  the  Higgs-boson searches background is discussed in Section \ref{Verification}.
     
\section{$H \rightarrow 4l$ background sources}
\label{background}
 
 The dominant {\it irreducible background} contribution to the Higgs boson searches in  the 
 $H \rightarrow ZZ^* \rightarrow 4l$ decay channel arises from the  direct production of 
 ($Z_1, Z_2$) pairs \cite{ATLAS_Higgs, CMS_Higgs}. Throughout this paper we shall use the 
 term $Z_1$  for the opposite sign and same flavour 
lepton pair with its invariant mass  higher  or equal $50$ GeV and the term  $Z_2$ for the pairs with 
in the remaining mass region. 
These terms are in close relation to those defined in \cite{CMS_Higgs}, where they represent, respectively, 
higher and lower invariant mass pairs.
In the  $m_{4l}$ region where the Higgs signal was reported  both definitions  can be considered as equivalent 
because  the remaining contributions of the ($Z_1,Z_1$) and ($Z_2,Z_2$) pairs  to four-lepton events 
are negligible\footnote{No  ($Z_2,Z_2$) events
and one ($Z_1,Z_1$) event have been found  in the mass region $120$--$130$ GeV in full the samples of 
the ATLAS and CMS collected data \cite{ATLAS_Higgs,CMS_Higgs}.}.

At the LHC the ($Z_1,Z_2$) pairs are produced in quark--antiquark annihilations, quark--gluon scattering and  
gluon--gluon fusions. In this paper we focus our attention on the first two contributions.   

The amplitude level mechanism of the  ($Z_1, Z_2$) pair production  can be represented  
as the Drell--Yan production of $Z/\gamma ^*$  accompanied by the initial or final state radiation of $\gamma ^*$, 
followed by its  conversion into the opposite charge and same flavour lepton pair%
\footnote{There can be also radiation/conversion of $Z^*$, but its contribution to the $Z_2$ mass region is negligible.}. 
The off-shell photons can be radiated both by the final state leptons coming from the $Z$-boson decays, 
and by the initial and (in the case of the quark--gluon processes) final state quarks. 
At the cross-section level, their origin cannot be unambiguously identified and the corresponding 
interference terms as well as the interference terms 
arising, in the case of the 4$\mu$ and 4$e$ final states, from the presence of indistinguishable fermions 
must, in principle, be taken into account. 
However, for the Higgs boson mass region discussed in this paper, the interference 
effects are expected to be small \cite{POWHEG}. 

In the ATLAS and CMS papers \cite{ATLAS_Higgs, CMS_Higgs} the irreducible background to the 
$H \rightarrow 4l$ decay channel is claimed to be controlled, for the quark--antiquark annihilation contribution, 
within $3$--$10$\% accuracy. This uncertainty was estimated by varying the factorisation and renormalisation scales, 
and by varying PDFs within their uncertainty range. 
The  missing contribution of the quark--gluon scattering processes  and the missing PS contribution were not accounted for
in the quoted above accuracy estimation.  

The event generator which was used in \cite{ATLAS_Higgs, CMS_Higgs} for the determination 
of the quark--antiquark annihilation background and its uncertainty 
was {\sf POWHEG} \cite{POWHEG}. This ME generator does not cover entirely the  phase space 
of virtual photon emissions, in particular 
the region populated by virtual photons emitted at the ``early stage"  of the PS development. 
As long as the region of $m_{Z_2} < 20$ GeV is avoided in the Higgs searches, such an 
approximation is claimed to be justified \cite{POWHEG}. For the searches reported 
in \cite{ATLAS_Higgs,CMS_Higgs}, which open the phase space towards smaller masses, 
the validity of such a statement remains to be verified.

The effect of the two approximations made in \cite{ATLAS_Higgs, CMS_Higgs}: (1) the missing contribution 
of the quark--gluon processes and (2) the missing PS contribution, even if expected to be small,
and consequently not taken into account in the ATLAS and CMS analyses, needs to be 
estimated to gain a full confidence in the claimed precision of the irreducible background level.

It remains to be mentioned that the contribution of the quark--gluon scattering  processes 
in which virtual photons are emitted in PS initiated by the final-state recoil quark may also
influence the size of the {\it reducible background} contribution. 
The  reducible background \cite{ATLAS_Higgs, CMS_Higgs} is dominated by $Z + jets$  events
(mostly $Z b \bbar$ and $t \tbar$) which give rise to the detector reconstructed objects 
faking the isolated lepton identification signatures. 
This background  is  detector and analysis method dependent. 
Its size was estimated using data driven methods \cite{ATLAS_Higgs, CMS_Higgs}. 
In general, these methods define control regions to monitor the background level and 
subsequently extrapolate its magnitude to the signal region, based on the assumption 
that the  $Z b \bbar$ and $t \tbar$ events are the sole background sources.

The processes of radiation of virtual photons by the outgoing quarks  
(so far not taken into account in the ATLAS and CMS analyses) produce isolated
leptons already in  the ``$Z$ plus  a single light-flavour jet'' events, as long as $m_{Z_2}$ is large enough 
to sweep out the leptons outside of the recoil quark jet cone. 
Inclusion of such a process 
could influence the shapes 
of the lepton track's ``Distance of the Closest Approach" (DCA) and 
the lepton candidate isolation criteria, leading to an underestimation 
of the reducible background level.  This effects is  expected
to be significantly amplified  in  the kinematical region selected for the Higgs bosons searches 
in which $m_{Z_2} > m_b$,  where $m_b$ is the mass of the $b$-quark. 
It has to be mentioned that the processes of radiation of large mass virtual photons are 
not generated by  the standard versions of the PS generators used by the LHC experiments.

\section{{\sf PYTHIA} precision}
\label{calibration} 

The fully inclusive $Z/ \gamma^*$ processes, including both the quark--antiquark 
and quark--gluon collisions,  were generated for 8 TeV proton--proton 
collisions using {\sf PYTHIA}~8.180. Only the leptonic decays were retained. The minimal 
mass of the $Z/ \gamma^*$ boson was set to be $50$ GeV. The minimal transverse momentum 
cut-off of virtual photons was equalised for the quark and lepton emissions to be $0.1$ GeV for 
time-like showers and $0.01$ GeV for space-like showers. Both cuts are significantly lower 
than the invariant masses of the virtual photons studied in this paper, $m_{\gamma^*} > $~5 GeV. 
The phase space for the generated virtual photon masses up to the value of $m_{\gamma^*} =  50$ GeV was open.
Hadronisation processes, except for prompt decays, were switched off, and the default CTEQ~5L PDFs \cite{CTEQ} were used. 

For the calibration of the {\sf PYTHIA} generator precision in describing the processes 
of large-mass virtual photon emissions by parton showers, we analyse first its performance  for the  
processes of $Z$-boson decays into four leptons, $Z \rightarrow 2l + \gamma^* \rightarrow 4l$. Technically,  
we select the region of $80\,{\rm GeV} < m_{4l} < 100\,{\rm GeV}$ where, for 
$m_{\gamma^*} > $~5 GeV,  a large majority (more than $97\%$) of 
virtual photons are emitted by the leptons coming from the $Z$-boson decays. 
This {\it calibration} region is particularly well suited for testing how precisely the {\sf PYTHIA} PS recoil
model mimics the exact  ME calculations. 

Throughout this paper the term lepton represents  the {\it dressed}\ lepton. 
For {\sf PYTHIA}, where the PS emissions of virtual photons 
are interleaved with the emissions of on-shell  photons, the origin of the 
latter was traced back to the mother leptons and their four-momenta were 
added to the four-momenta of the mother leptons to form dressed leptons\footnote{The inclusion 
of virtual photons in the dressing procedure can be neglected for the specified above 
cut-off values of the minimal transverse momentum of the virtual photons.}. 

The departure point of our analysis is the comparison the shapes of the distributions obtained with 
{\sf PYTHIA} and {\sf POWHEG}.

The $ZZ$ pair production events of {\sf POWHEG-BOX} version 2129 were  generated and 
analysed  in our studies. We took into account  the full set of the NLO ME processes
leading to the production of four charged leptons,   including both the 
quark--antiquark and quark--gluon collisions. In order to study the 
magnitude of the interference effects two samples of events were generated, 
including and excluding the interference effects. 
For the minimum mass of the lepton pairs coming from the $Z/\gamma^*$ decays 
we required $m_{ll} > 5\,$GeV. Since in {\sf POWHEG} the QED radiation processes are not 
taken into account,  its final-state leptons are, by definition, 
the {\em dressed}\ leptons. The {\sf MSTW2008nlo68cl} parametrisation \cite{MSTW} of the PDFs 
was used in the event generation.

In this paper,  {\sf PYTHIA} and {\sf POWHEG} event generation is restricted 
to the  $4\mu$ final state. Their extension to an  arbitrary flavour mixture of the lepton pairs
is straightforward. 
The choice of the $4\mu$ final state maximises: (1)  the 
interference effects related to the presence of indistinguishable fermions 
in the final state, and (2) the effects caused by mis-association of the  leptons to their  respective 
$Z/\gamma^*$ and $\gamma ^*$ parents.  The magnitude of the interference and mis-association, 
determined with such a sub-sample of events,  provides an upper limit of these effects 
for an arbitrary flavour composition of the four-lepton final state.

\begin{figure}[h]
\begin{center}
\includegraphics[width=1.00\textwidth]{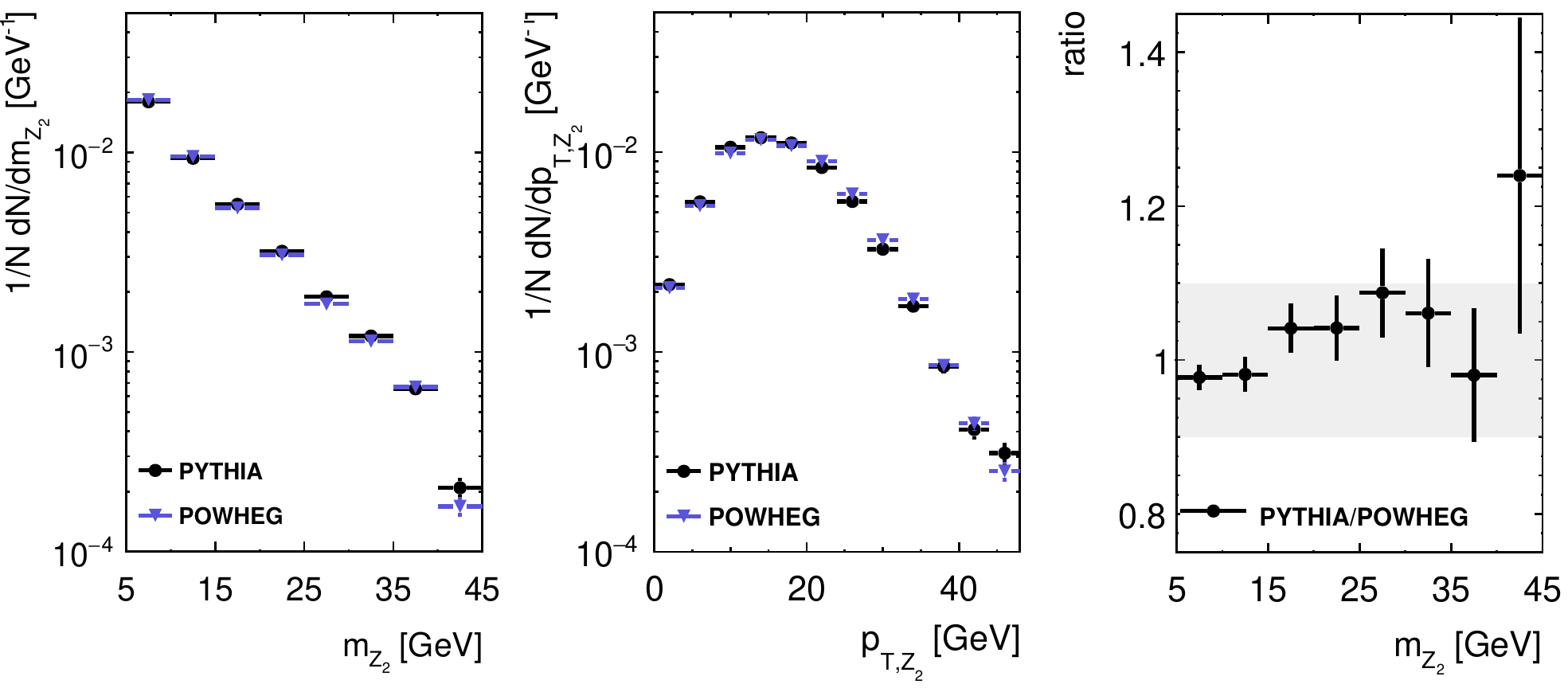} 
\end{center}
\caption{The comparison of the distributions of the invariant mass $m_{Z_2}$ (left)  and the transverse 
momentum   $p_{T,Z_2}$ (middle) of virtual photons in the $Z$-resonance region $80\,{\rm GeV} < m_{4l} <100\,{\rm GeV} $
for {\sf PYTHIA} and {\sf POWHEG}. The {\sf POWHEG} interference effects are neglected in this comparison. 
Only $4\mu$ events satisfying the: (1) $m_{Z_2} > 5$ GeV and (2) $p_{T,l} > 4$ GeV conditions contribute to these plots. 
The right plot shows the ratio of the  $m_{Z_2}$ distributions for  {\sf PYTHIA} and {\sf POWHEG}.}
\label{Resonance}
\end{figure}

In Fig.~\ref{Resonance} we show the distributions of the mass,  
$m_{Z_2}$, and  transverse momentum, $p_{T,Z_2}$,   of the  $Z_2$ lepton pairs for {\sf PYTHIA} and 
{\sf POWHEG} and the ratio of their   $m_{Z_2}$ distributions. 
The interference effects in {\sf POWHEG} are neglected for these plots. 
In the generated  sample of events the transverse momentum of each of the leptons satisfies 
the condition $p_{T,l} > 4$ GeV, representing roughly the lower limit of the ATLAS and CMS 
isolated lepton detection and reconstruction acceptance. 
In all the plots shown in Fig.~\ref{Resonance},  $Z_2$ represents the pair of the 
opposite charge leptons remaining after selection of  the leading $Z_1$ pair.
The leading $Z_1$ pair is chosen as the one having  its reconstructed mass closest to the $Z$-boson mass.

We find an unexpectedly good agreement of the {\sf PYTHIA} and 
{\sf POWHEG} distributions  over  the full  range of the $Z_2$ pairs masses
and over the full range of their transverse momenta.
It has to be reminded here that {\sf PYTHIA} uses the parton shower approximations 
in the extended virtual photon phase-space region,  up to the invariant mass of of $50$ GeV -- 
the region where such approximations  have never been tested. 

In  general, and particularly in the $Z$-resonance region of $80\,{\rm GeV} < m_{4l} <100\,{\rm GeV} $, 
the association of the leptons to the $Z_1$ and $Z_2$ pairs may not reflect their amplitude-level association to the $Z/\gamma^*$ and $\gamma^*$.
Assuming, for a while,  the absence of the interference terms,  the effect of such an  algorithmic miss-association can be determined using  the 
{\sf PYTHIA} sample of events because there each lepton can be traced back to its mother particle.  
\begin{figure}[htb!]
\begin{center}
\includegraphics[width=1.00\textwidth]{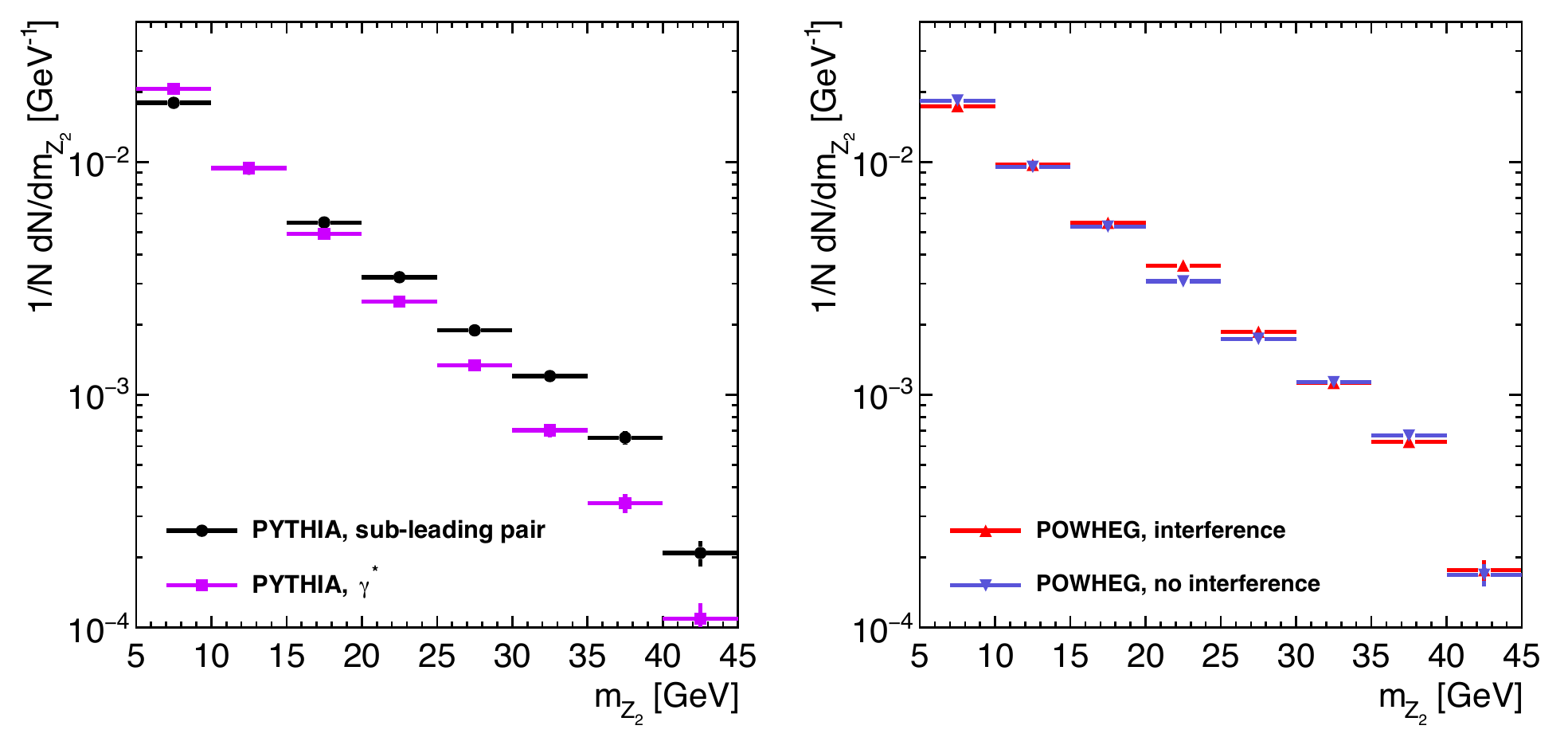} 
\end{center}
\caption{(Left) The comparison of the {\sf PYTHIA} distributions of the invariant mass $m_{Z_2}$:  
(1)  for the $Z_2$ pairs found algorithmically,   and (2) for $Z_2$  the pairs coming 
from conversions of virtual photon emitted by leptons and quarks. 
(Right) The comparison of the {\sf POWHEG} distributions of the invariant mass $m_{Z_2}$:  
(1) including and (2) excluding the interference effects.}
\label{effects}
\end{figure}
In the left panel of Fig.~\ref{effects} the resulting miss-association bias is represented as  the difference 
between the $m_{Z_2}$  distributions: (1)  for the $Z_2$ pairs found algorithmically  and (2) for the $Z_2$  pairs coming 
from conversions of virtual photons emitted by leptons and quarks.  

The  predicted {\sf PYTHIA}  distributions, prior to any comparison with the data, 
must be corrected for the missing interference effects.
The corrections are determined using  {\sf POWHEG} 
by taking the ratio of the distributions obtained by including (excluding)
the interference effects. 
The size of the interference corrections to the $m_{Z_2}$  distribution
is illustrated in the right panel of Fig.~\ref{effects} where {\sf POWHEG} 
distributions are shown twice, including and excluding the interference effects.
Since the  interference  effects are absent for the $Z \rightarrow 2\mu 2e$  decays,  
the plotted differences for the $Z \rightarrow 4\mu$ channel are larger than those  
for  the all-flavour inclusive background to Higgs searches.   

It is interesting to note that the miss-association effects are significantly larger than the 
interference effects. They have never been studied before because   the logic 
of the {\sf POWHEG} generator forbids tracking back the final leptons to their amplitude-level 
origin. 
The size of the miss-association effects underlines the necessity of using precisely the same 
pair-association 
algorithm for  data and for Monte Carlo  samples of events. In the following we shall 
use only the algorithmic association of leptons to the  $Z_1$ and $Z_2$ pairs, defined in  
Section~\ref{calibration}.

Since {\sf PYTHIA} is a LO-type generator, it is necessary to apply a $K$-factor to match 
the absolute normalization of cross-sections.  We exploit the ATLAS $Z \rightarrow 4l$ 
data \cite{ATLAS_Z4l} to determine the $K$-factor for subsequent studies. The ATLAS experiment 
measured the total $Z \rightarrow 4l$ cross-section in the $Z$-boson resonance peak,  $80\,{\rm GeV} < m_{4l} <100\,{\rm GeV} $, 
requiring the minimal mass of muon or electron pairs $m_{l^+ l^-} > 5$ GeV, to be
$\sigma^{\rm data}_{Z \rightarrow 4l} = 107 \pm 9\, (\rm stat.) \pm 4\, (\rm syst.)  \pm 3\, (\rm lumi.)$ fb. 
The corresponding cross section calculated with {\sf PYTHIA}  is $\sigma^{\sf PYTHIA:LO}_{Z \rightarrow 4l} = 89$  fb. 
In the following we shall thus apply the $K$-factor of $1.2$ to all the cross-section calculations. 

Assuming {\sf POWHEG} to be the precision template for our studies, 
we conclude the discussion presented in this section by the statement 
that the processes of emission of virtual photons from the outgoing leptons are controlled 
by the {\sf PYTHIA} generator with a precision better than $\sim 10\%$.  
This defines the precision level of the studies presented in the next section.

\section{Background calculation results}
 \label{results}

Having analyzed  the precision of the {\sf PYTHIA}  model of the emission of large-mass virtual photons from the 
final state leptons,  we focus now  our attention on  the  processes in which the virtual photons are emitted 
not only by the final-state leptons but,  predominantly,  by  the the PS processes initiated by quarks and gluons. 

The extension  of  the ``leptonic  calibration"  of the  {\sf PYTHIA} precision  to quark radiation processes is based 
on the assumption that  the strength of the QCD  confinement forces and the precise values of the quark masses,
which, in principle, may lead to significant differences in photon emissions by leptons and quarks,  become  
irrelevant  for  the process of highly virtual,   $m_{\gamma ^*} \gg m_e, m_q, \Lambda _{QCD}$,  photon emissions. 
Assuming its validity, quarks, except for the differences in the 
electric charges,  can be considered as equivalent emitters of highly virtual photons  as leptons. Consequently, 
the control of  the {\sf PYTHIA} precision,  estimated using virtual photon emission by leptons, can be extended 
to processes involving quarks.

In the four-lepton invariant mass region of $100\,{\rm GeV} < m_{4l} < 2m_Z $
the dominant process producing four-lepton final state is the radiation 
of virtual photons from the initial and final state quarks involved in the Drell--Yan 
production of $Z/\gamma ^*$. The processes of leptonic radiation 
of virtual photons contribute at  $<10\%$ level, populating mainly in the region of masses
close to the lower boundary of this mass region\footnote{This small admixture of 
the leptonic radiation events arises from the Breit--Wigner tail of the $Z$-resonance.}.

The results presented in this section are based on the {\sf PYTHIA} sample of 
$Z/\gamma ^*$ Drell--Yan events generated in the region of $ m_{Z/\gamma ^*} >50$  GeV. 
Both leptons and (PS) quarks participating in the $Z/ \gamma^*$-boson production process 
are allowed to radiate virtual photons, provided that their  invariant mass is  below $50$ GeV. 
For the virtual photon radiation from the quarks  the $Z_1$ boson 
is thus {\it always} associated with the $Z/ \gamma ^*$-boson matrix element, while the $Z_2$ boson 
is {\it always}  generated by the {\sf PYTHIA} parton showers.
To take into account a reverse and symmetric configuration  which is not generated, 
all events where the origin of $Z_2$ is traced back to a quark acquire the weight
equal to $2$. A fraction of the  $4l$ phase space
remains, however,  uncovered in such a simplified event generation procedure. The missing events  
not taken into account in the  {\sf PYTHIA} generation process,  which may potentially 
contribute to the background to the Higgs searches,  contain the  ($Z_1,Z_1$) 
and ($Z_2,Z_2$) pairs.  Their contribution  to the background 
under the Higgs peak, $120\,{\rm GeV}   <m_{4l} < 130\,{\rm GeV} $,  calculated using  {\sf POWHEG},
is below $1\%$. In the extended region,  $100\,{\rm GeV}  < m_{4l} < 160\,{\rm GeV}$,  investigated in this section, 
their integrated contribution is below $2.5\%$ and is peaking,  respectively for the  ($Z_1,Z_1$) 
and ($Z_2,Z_2$) pairs, in the regions close to the low and high $m_{4l}$ region boundaries. 
The above missing phase-space contributions can be thus safely neglected in the studies presented 
in this paper. 

In the following,  we compare the predictions of {\sf PYTHIA} and {\sf POWHEG} 
for the Higgs searches background in the $4\mu$ channel as a function of the phase-space 
cut on the  kinematical variable which, as discussed in more details in \cite{DDYP1}, 
drives the background peak position in the $m_{4l}$ distribution: the minimal
allowed transverse momentum of each of the four leptons,  $p_{T,l}$. 
For this comparison,  the {\sf PYTHIA} distributions were  corrected for the missing interference effects 
using the ratios of the  {\sf POWHEG}  distributions obtained by including  and excluding  
the interference effects in the event generation.

\begin{figure}[h]
\begin{center}
\includegraphics[width=1.00\textwidth]{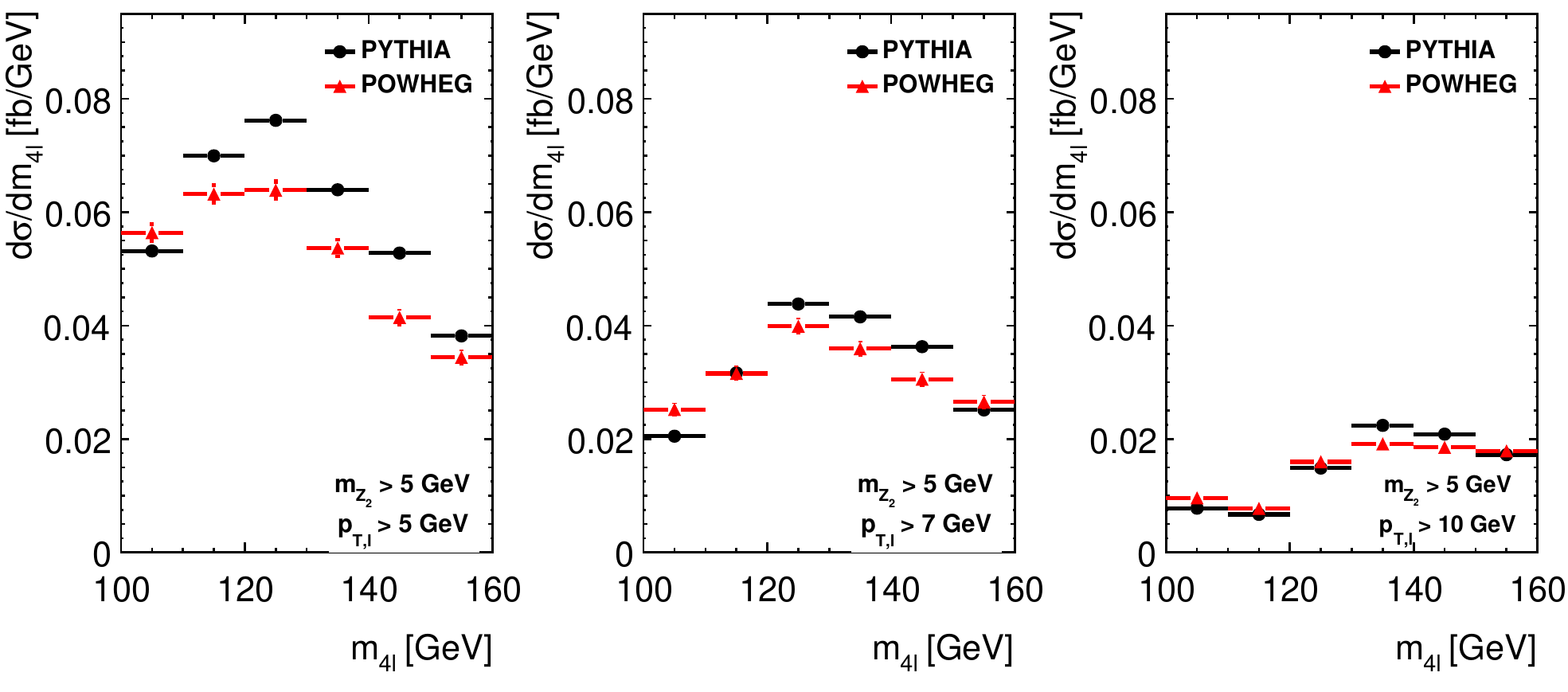} 
\end{center}
\caption{The $4\mu$  channel {\sf PYTHIA} and {\sf POWHEG} differential cross-sections 
as a function of $m_{4l}$ for $m_{Z_2} > 5$ GeV 
and the minimal $p_{T,l}$ cut values: (left) $5$ GeV, (middle) $7$ GeV, and (right) $10$ GeV.
The {\sf PYTHIA} cross-sections are  corrected for the missing interference effects 
using the ratios of the  {\sf POWHEG}  distributions obtained by including  and excluding  
the interference effects in the event generation. }
\label{cuts1}
\end{figure}

In Fig.~\ref{cuts1} we show the differential cross section as a function of 
$m_{4l}$ for the sample of the $4\mu$
events  for which $m_{Z_2} > 5$ GeV for the following three $p_{T, l}$ cuts: 
$5$, $7$, $10$ GeV, and compare the results of the two event generators. 
We observe a satisfactory agreement within the accuracy of the presented studies of $10\%$ for 
$p_{T,l}$ cut values of $7$ GeV and  $10$ GeV, while
for the $p_{T,l}$ cut value of $5$ GeV the background in the Higgs signal region is $20\%$ higher
for  the {\sf PYTHIA} predictions compared to the {\sf POWHEG} ones. 
As it is the case in DDYP \cite{DDYP1}, 
the mass distribution peak position varies with the 
increasing $p_{T, l}$ cut.
It is interesting to point out that the background level in the Higgs peak region, 
calculated  with {\sf PYTHIA},  varies more rapidly as a function of the  $p_{T, l}$ cut value
compared to {\sf POWHEG}. 

We conclude that  the effects of including the PS 
processes producing virtual photons do not 
appear to change significantly the Higgs searches background level, 
provided that the $p_{T, l} < 7$ GeV region is avoided.  Such a phase-space restriction  is justified 
if the excess events in the $m_{4l} \sim 125$ GeV region are 
assumed to originate from the  Higgs boson decays, producing 
rarely leptons carrying such a small transverse momentum. 
However, for the experimental investigation of less theoretically biased scenarios 
the  $p_{T, l}$ cut should be lowered as much as it is experimentally possible. In  such a case 
the PS effects will have  to be taken into account.

\section{Testing experimentally importance of PS}
\label{Verification}

For the canonical Higgs searches cuts 
applied in the ATLAS and CMS analyses \cite{ATLAS_Higgs,CMS_Higgs}  the PS effects
calculated using the LO {\sf PYTHIA} generator can be neglected. 
However, if the phase space for the emission of virtual photons is open not only 
towards smaller values of  $p_{T, l}$ but also towards smaller 
values of  $m_{Z_2}$, our studies indicate that they  become important 
--  leading to a stronger increase of the Higgs searches background with decreasing  $m_{Z_2}$ 
than predicted by {\sf POWHEG}. 

\begin{figure}[h]
\begin{center}
\includegraphics[width=1.00\textwidth]{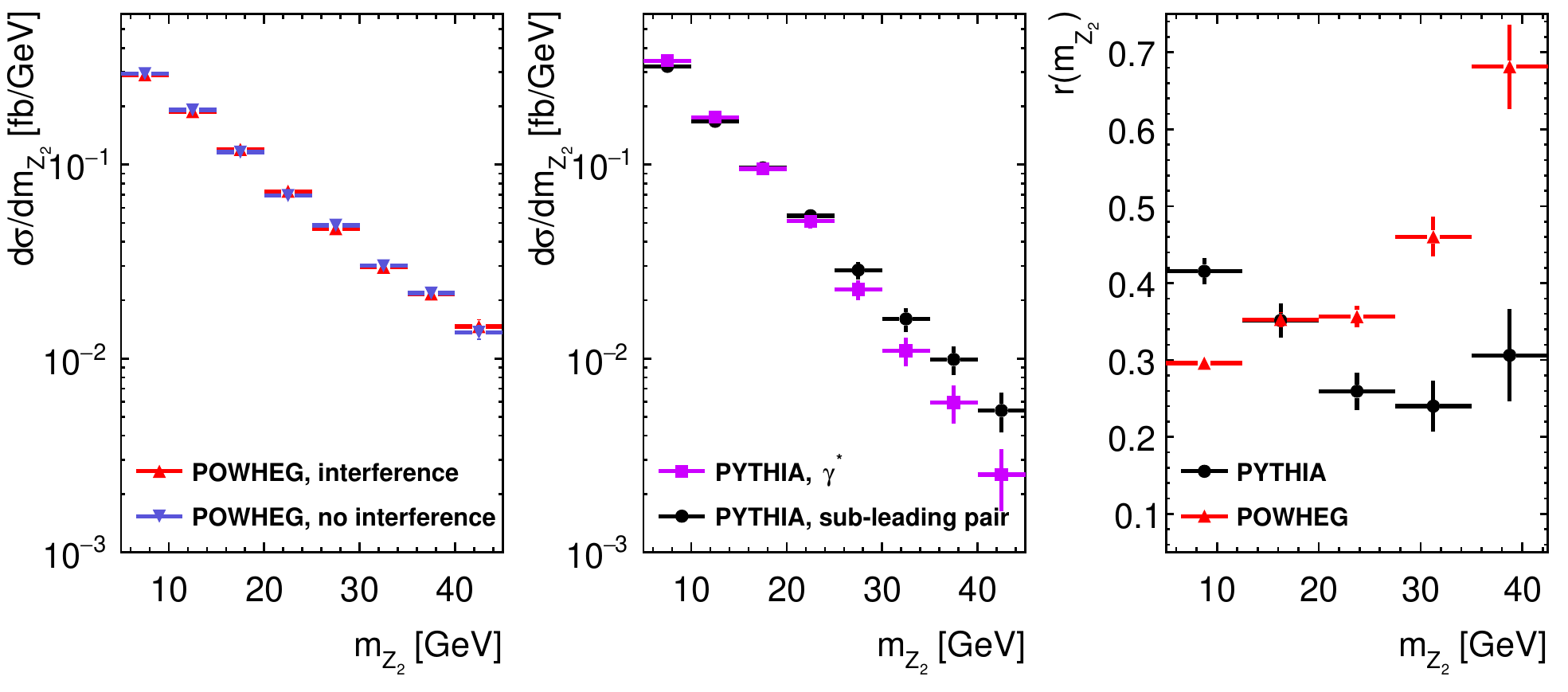} 
\end{center}
\caption{The comparison of the  {\sf PYTHIA} and {\sf POWHEG} differential cross-sections as a function of 
$m_{Z_2}$ for the  $4\mu$ event sample restricted by the following cuts: 
$100\,{\rm GeV}  < m_{4l} < 160\,{\rm GeV}$ and $p_{T,\mu} > 4 $ GeV. The interference and the mis-assignment 
effects are shown in the left and central panel, respectively. 
The $r(m_{Z_2})$ ratio, defined in the text, is shown in the right panel for the  $4\mu$ event sample restricted by the cuts
$p_{T,\mu} > 4 $ GeV and $m_{Z_2} > 5$ GeV.}
\label{Plot4}
 \end{figure}

This is illustrated in the left and central panels of  Fig.~\ref{Plot4}, showing the differential cross-sections as a function of 
$m_{Z_2}$ for the $4\mu$ event selection satisfying the  
$100\,{\rm GeV}  < m_{4l} < 160\,{\rm GeV}$, $p_{T,\mu} > 4 $ GeV and $m_{Z_2} > 5$ GeV conditions. Indeed, 
the mass spectrum is steeper in {\sf PYTHIA} than in {\sf POWHEG}. The interference and the 
mis-assignment effects  are significantly smaller in the selected $m_{4l}$ mass region 
compared to the $Z \rightarrow 4l$ region, indicating that the spectra of virtual photon masses are sufficiently distinct 
for the two  generators -- to be resolved experimentally. 

The effect of the missing PS contribution for low values of $m_{Z_2}$  
 may have already been observed in the CMS analysis of 
the contributions to the $H \rightarrow W^+W^-$ background coming from 
the $W \gamma ^*$  production processes \cite{Kilic, CMS_H_WW} --
more events were observed than predicted using exclusive ME calculations. 
This effect was absorbed in \cite{CMS_H_WW} within the large value and the large uncertainty 
of the estimated $K$-factor of $1.5 \pm 0.5$. Such a  $K$-factor was necessary 
to  rescale upwards the {\sf MAGDRAPH} matrix element \cite{MADGRAPH} calculations to 
match the observed  $m_{\mu^+\mu^-}$ spectra in the mass region below $12$ GeV 
for $l^{\pm} \mu ^+ \mu^-$ events\footnote{Note,  that $W \gamma ^*$
processes may contribute as background to the Higgs
boson signal whenever one of the three leptons in the final state is not selected, 
no matter what is the  mass of the lepton pairs, provided that only one of the two 
leptons  from internal $\gamma ^*$ conversion is emitted 
at sufficiently large transverse momentum.}. 

Concluding   the studies presented in this paper 
we  propose to measure the following precision observable 
which is particularly sensitive to the  the parton shower effects:  
\begin{eqnarray}
r(m_{Z_2}) = \frac{d \sigma/dm_{Z_2} ( 100\, {\rm GeV} <  m_{4l} < 160\,{\rm GeV})}
{d \sigma/dm_{Z_2} ( 80\, {\rm  GeV} <  m_{4l} < 100\,{\rm GeV})}. \,
\label{eq:rmZ2}
\end{eqnarray}

This ratio is  experimentally robust, i.e. insensitive to a large fraction 
of the systematic measurement  error sources (they cancel in the ratio).
In addition, it is robust with respect to the approximations
inherent to the modelling of virtual photon emission in PS.  
In our view, measurements of such a ratio, preferentially at two LHC 
collision energies: $8$ TeV and $13$ TeV, could restrict experimentally 
the size of the PS effects and assure more robust predictions for the Higgs searches
background, with the precision significantly higher than that of the present studies. 
 
In Fig.~\ref{Plot4} (right panel) we show the predictions of both event generators 
for the $r(m_{Z_2})$ ratio, extended down to the values of  $m_{Z_2}$ used in the 
studies of the $Z \rightarrow 4l$ decays in \cite{ATLAS_Z4l}.
This ratio was calculated for the $4\mu$ events, with each of the 4 muons satisfying 
the conditions of $p_{T,\mu} > 4 $ GeV and $m_{Z_2} > 5$ GeV. 
The {\sf PYTHIA} value of $r(m_{Z_2})$ was corrected in this plot for the missing interference effects 
using the ratios of the  {\sf POWHEG}  distributions obtained by including  and excluding  
the interference effects in the event generation. 

We see a clear difference between the predictions of the two event generators.  
They  may be resolved  experimentally  
using  the collected $7$ and  $8$ TeV data,
provided that the analysis is extended to the $m_{Z_2}$ mass region 
below $12$ GeV, where the $30$--$40\%$ excess of the 
data with respect to the {\sf POWHEG} background  is predicted\footnote{The 
difference of the  {\sf PYTHIA} and {\sf POWHEG} predictions is more pronounced  
in the $m_{Z_2} > 30$ GeV mass region. However, 
the number of events in this region collected in the 7 and 8 TeV LHC runs   
is too small to provide  a statistically conclusive test.}.

Even if the extension of the $H \rightarrow ZZ^*$ acceptance cuts 
towards  lower $m_{Z_2}$  (and lower $p_{T,l}$) 
would not increase significantly the acceptance for the Higgs decay events, 
it could replace the belief in the adequacy of the presently 
available theoretical  tools by the confidence in experimentally
understanding the Higgs searches background sources.

\section{Conclusions}
\label{Conclusions}

The ATLAS and CMS analyses of the irreducible background to the Higgs-boson searches in the 
four-lepton channels presented in Refs.~\cite{ATLAS_Higgs,CMS_Higgs} leave, in our opinion, 
two open questions: 
\begin{enumerate}  
\item  
Can  the contribution to the Higgs searches background coming from the quark--gluon scattering processes be neglected?
\item 
Can the contribution of  the processes 
of high-mass virtual photons emissions by the initial and final state parton showers (PS)  be 
neglected? 
\end{enumerate} 
In this paper we have addressed the second question,
while the first one will be  investigated in detail in a separate publication \cite{Anatoli_future}. 

Within the precision  inherent to the LO-type {\sf PYTHIA} generator,  
``calibrated"  using the processes of virtual photon emissions 
by leptons to $\sim 10 \%$,  the answer to the second question is affirmative. These processes
indeed can be neglected at such a precision level for the  phase-space cuts applied in 
the Higgs boson targeted  searches \cite{ATLAS_Higgs,CMS_Higgs}.
 
A measurement method tailored for pinning down the PS effects,  
of particular importance for searches of alternative/complementary mechanisms 
producing the excess of events in the 125 GeV mass region, 
has been proposed.
This method allows the PS  effects to be established experimentally 
in the PS-sensitive $m_{Z_2}<12 $ GeV mass region,  
where we predict the $30$--$40\%$ excess of 
data with respect to the {\sf POWHEG} background
for the $8$ TeV sample of the $4l$ events collected at the LHC.  
The sensitivity to the PS effects is expected to increase in the 
the subsequent phase of the LHC operation and may become 
important already for the canonical phase-space cuts.  

It would be advantageous if the corresponding experimental studies 
could  be supported  on the theoretical side  by constructing 
a MC generator for the four-lepton production processes 
in which,  on top of the {\sf POWHEG} NLO QCD
ME calculations, the interleaved NLO QCD and QED parton showers (including 
virtual photon emissions) are incorporated.


\begin{thebibliography}{00}


\bibitem{ATLAS_Higgs}
ATLAS Collaboration,  Phys.\ Lett.\ B\ {\bf 716} (2012) 1. \\
ATLAS Collaboration,  Phys.\ Lett.\ B\ {\bf 726} (2013) 88. \\
ATLAS Collaboration, Phys.\ Rev.\ D\ {\bf 90},  (2014) 052004. \\
ATLAS Collaboration CERN-PH-EP-2014-170, to appear in Phys.\ Rev.\ D.
\bibitem{CMS_Higgs}
CMS Collaboration, Phys.\ Lett.\ B {\bf 716} (2012) 30. \\
CMS Collaboration, Phys.\ Rev.\ D {\bf 89} (2014) 92007.
\bibitem{DDYP1}
M.~W. Krasny and W. P{\l}aczek, Acta Phys.\ Pol.\ B\ {\bf 45} (2014) 71.
\bibitem{DDYP2}
M.~W. Krasny and W. P{\l}aczek,  arXiv:1501.04569 [hep-ph], to appear in Acta Phys.\ Pol.\ B.
\bibitem{KRASNY_S_dependence}
M.~W. Krasny,  Acta Phys.\ Pol.\ B\ {\bf 42} (2011) 2133.
\bibitem{POWHEG}
T. Melia, P. Nason, R. Rontsch, G. Zanderighi, JHEP {\bf1111} (2011) 78.\\
P. Nason, JHEP {\bf411} (2004) 040. \\
S. Frixione, P. Nason and C. Oleari, JHEP {\bf711} (2007) 70. \\
S. Alioli, P. Nason, C. Oleari and E. Re, JHEP {\bf1006} (2010) 43.
\bibitem{BvNB:1988}
F.~A. Berends, W.~L. van Neerven, G. Burgers, Nucl.\ Phys.\ B {\bf 297} (1988) 429.
\bibitem{Anatoli_future}
A. Fedynich, M.~W. Krasny and W. P{\l}aczek,  in preparation, to appear in Acta Phys.\ Pol. B.
\bibitem{PYTHIA}
T. Sjostrand, S. Mrenna, and P. Z. Skands, Comput.\ Phys.\ Commun.\ {\bf 178} (2008) 852, arXiv:0710.3820 [hep-ph].
\bibitem{CTEQ}
Hung-Liang Lai et al., Phys.\ Rev. D\ {\bf 82} (2010) 074024.
\bibitem{MSTW}
A.D. Martin, W.J. Stirling, R.S. Thorne, G. Watt, Eur.\ Phys.\ J.\ C {\bf 63} (2009) 189. 
\bibitem{ATLAS_Z4l}
ATLAS Collaboration, Phys.\ Rev.\ Lett.\ {\bf 112} (2014)  231806.
\bibitem{ATLAS_Z}
ATLAS Collaboration, Phys.\ Rev.\ D\ {\bf 85}  (2012) 72004.
\bibitem{Kilic}
Richard C. Gray, Can Kilic, Michael Park, Sunil Somalwar and  Scott Thomas, arXiv:1110.1368,  [hep-ph]. 
\bibitem{CMS_H_WW}
CMS Collaboration, JHEP {\bf 1401} (2014) 96.
\bibitem{MADGRAPH}
J. Alwall et al., JHEP {\bf 6} (2011) 128.
\end{thebibliography}
\end{document}